\documentclass[a4paper]{spie}
\usepackage{amsmath,amsfonts,amssymb}
\usepackage{graphicx}

\title{Discrimination of discord in separable Gaussian states}

\author{Gaetana Spedalieri}
\affil{Research Laboratory of Electronics, Massachusetts Institute
of Technology, Cambridge, Massachusetts 02139, USA \& Computer
Science, University of York, York YO10 5GH, UK}
\author{Stefano Pirandola, Samuel L. Braunstein}
\affil{Computer Science, University of York, York YO10 5GH, UK}

\begin{document}
\maketitle

%%%%%%%%%%%%%%%%%%%%%%%%%%%%%%%%%%%%%%%%%%%%%%%%%%%%%%%%%%%%%%%%%%%%%%%% OK

\begin{abstract}
Consider two bosonic modes which are prepared in one of two
possible Gaussian states with the same local energy: either a
tensor-product thermal state (with zero correlations) or a
separable Gaussian state with maximal correlations (with both
classical and quantum correlations, the latter being quantified by
quantum discord). For the discrimination of these states, we
compare the optimal joint coherent measurement with the best local
measurement based on single-mode Gaussian detections. We show how
the coherent measurement always strictly outperforms the local
detection strategy for both single- and multi-copy discrimination.
This means that using local Gaussian measurements (assisted by
classical communication) is strictly suboptimal in detecting
discord. A better performance may only be achieved by either using
non Gaussian measurements (non linear optics) or coherent
non-local measurements.
\end{abstract}


\section{Introduction}

Quantum correlations are today recognized as a fundamental
physical resource in quantum
information~\cite{Nielse,first,Hayashi,RMP,HolevoBOOK}. For pure
states of a quantum system, quantum correlations are exactly the
same as entanglement. However, this equivalence fails when general
mixed states are taken into account: Separable mixed states can
still have correlations which cannot be simulated by any classical
probability distribution~\cite{Qdiscord,Qdiscord2,VedralRMP}.
These correlations are quantified by the concept of quantum
discord, whose definition is related to inequivalent extensions of
the mutual information from the classical to the quantum setting.

On the one hand, the mutual information $I(X:Y)$ between two classical
variables, $X$ and $Y$, can be extended to the quantum mutual information
between two quantum systems, $A$ and $B$, defined by
\begin{equation}
I_{tot}(A:B)=S(A)+S(B)-S(A,B)~,
\end{equation}
where $S(.)$ is the von Neumann entropy. This quantity accounts for all the
correlations between the two quantum systems. On the other hand, another
possible extension of $I(X:Y)$ is given by the entropic quantity%
\begin{equation}
C(A:B):=S(A)-\inf_{\mathcal{B}}S_{\mathcal{B}}(A|B)~, \label{Cdef}%
\end{equation}
where $S_{\mathcal{B}}(A|B)$ is the conditional entropy of system $A$ given
system $B$ being subject to a quantum measurement $\mathcal{B}$, i.e., a
positive operator valued measure (POVM).

The quantity $C(A:B)$ is generally less than or equal to
$I_{tot}(A:B)$, and can be interpreted as the maximal amount of
classical correlations between $A$ and $B$. Quantum discord is
therefore defined as the difference between total and classical
correlations, i.e.,
\begin{align}
D(A  &  :B):=I_{tot}(A:B)-C(A:B) =S(B)-S(A,B)+\inf_{\mathcal{B}}S_{\mathcal{B}}(A|B)~. \label{Ddef}%
\end{align}
In general, $C$ and $D$ are asymmetric quantities under system permutation
$A\leftrightarrow B$, unless we consider symmetric quantum states, i.e., such
that $\rho_{AB}=\rho_{BA}$.

In the continuous variable framework, systems $A$ and $B$ are typically
bosonic modes in a Gaussian state. In this case, we may restrict the
minimization in Eq.~(\ref{Ddef}) from arbitrary measurements $\mathcal{B}$ to
Gaussian measurements $\mathcal{G}$ and define the Gaussian
discord~\cite{GerryPRL,MatteoPRL}. Gaussian discord can be easily computed and
it has been proven to be the actual (unrestricted) quantum discord for a large
family of Gaussian states~\cite{OptimalityDiscord}. In particular, this
equivalence is true for the Gaussian states considered here.

In this paper, we study the performance of global and local
detectors in discriminating the presence of correlations in
bipartite Gaussian states. More precisely, we compare an optimal
coherent detector with local Gaussian detectors in the
discrimination of quantum discord and classical correlations,
assuming either a tensor-product of single-mode thermal states or
a correlated (but separable) two-mode thermal state. We quantify
the advantage of the coherent detector both in the setting of
single-copy discrimination and that of multi-copy discrimination
(by comparing the error-exponents in the decaying error
probability). Because this advantage is strictly larger than zero,
one may hide classical information in the separable correlations
of Gaussian states, therefore realizing a simple Gaussian form of
quantum data hiding~\cite{hiding}.

\section{Discrimination scenario}

Let us consider two bosonic modes, $A$ and $B$, prepared in a symmetric
Gaussian state $\rho$. This state is randomly chosen from a binary ensemble
$\{\rho_{0},\rho_{1}\}$ with uniform probability $p_{0}=p_{1}=1/2$. In other
words, a bit of information $k=0,1$ is encoded into the state of the two
bosonic modes. The two Gaussian states $\rho_{0}$ and $\rho_{1}$ are taken to
be separable but with maximal difference in terms of correlations. We consider
$\rho_{0}$ to be tensor product of two thermal states, so that it has zero
discord ($D_{0}=0$) and zero classical correlations ($C_{0}=0$). For $\rho
_{1}$, we consider a\ separable Gaussian state with the same energy as
$\rho_{0}$ but maximally correlated, i.e., with maximal discord $D_{1}$ and
maximal classical correlations $C_{1}$. In other words, for fixed energy, we
are comparing a completely uncorrelated Gaussian state with the most
correlated (but separable) Gaussian state, so that the bit of information is
practically encoded in the absence or presence of correlations with maximal
variations $\delta D:=D_{1}-D_{0}$ and $\delta C:=C_{1}-C_{0}$.

\begin{figure}[ptbh]
\vspace{-3.4cm}
\par
\begin{center}
\includegraphics[width=0.99\textwidth] {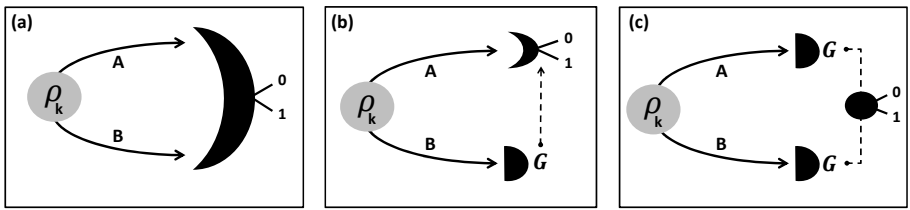}
\end{center}
\par
\vspace{-4.5cm}\caption{Discrimination of two Gaussian states,
$\rho_{0}$ and $\rho_{1}$, using a coherent detector (inset a), or
using an incoherent detector based on a local Gaussian measurement
followed by an Helstrom measurement (inset b), or using an
incoherent detector based on two local
Gaussian measurements whose outputs are suitably post-processed (inset c).}%
\label{pic1}%
\end{figure}

For state-discrimination, i.e., bit-decoding, we then consider two different
types of detection as depicted in Fig.~\ref{pic1}. First, we consider the
optimal global measurement on both modes $A$ and $B$. This is given by the
Helstrom POVM$\ \{E_{0},I-E_{0}\}$, with $E_{0}$ projecting on the positive
part of $\rho_{0}-\rho_{1}$, which is clearly a non-Gaussian measurement. The
minimum error probability is given by the Helstrom bound~\cite{Helstrom}%
\begin{equation}
P=\frac{1}{2}[1-T(\rho_{0},\rho_{1})]~,
\end{equation}
where $T$ is the trace distance between the two states. Correspondingly, the
information retrieved is equal to%
\begin{equation}
I=1-H(P)~,
\end{equation}
where $H$ is the binary Shannon entropy.

The second type of detection corresponds to local measurements (see
Fig.~\ref{pic1}). Here the decoder applies an optimal Gaussian measurement on
mode $B$, followed by an optimal Helstrom POVM on mode $A$. This local
measurement clearly represents an upper bound for all the possible local
Gaussian measurements (where both the modes are detected by optimized Gaussian
POVMs). This second scheme will be affected by an error probability which is
generally bigger than before, i.e., $P_{loc}\geq P$, with retrieved
information
\begin{equation}
I_{loc}=1-H(P_{loc})\leq I~.
\end{equation}
Here we are interested in studying the behaviour of these
quantities (probabilities and mutual informations) in terms of the
encoded correlations, i.e., the variation of Gaussian discord
$\delta D$ and the variation of classical correlations $\delta C$.
In particular, we are interested in investigating the behaviour of
the information gain $\Delta I:=I-I_{loc}\geq0$ of the global
detection versus the local one. We show that there is a strict
separation $\Delta I>0$ so that global detection is proven to
provide the best strategy in detecting such kinds of
correlations.\begin{figure}[ptbh] \vspace{-3.5cm}
\par
\begin{center}
\includegraphics[width=0.9\textwidth] {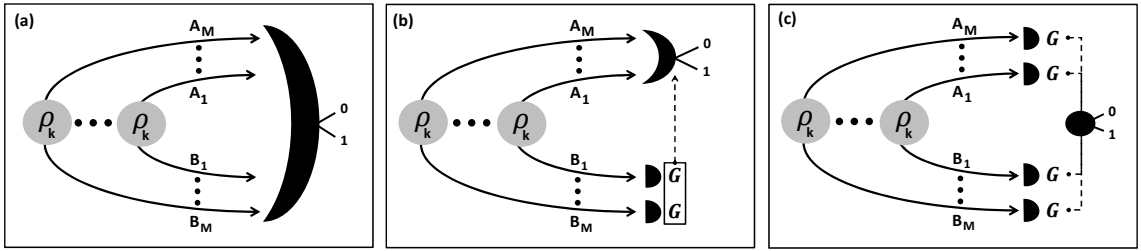}
\end{center}
\par
\vspace{-3.6cm}\caption{Multi-copy discrimination of two Gaussian
states, $\rho_{0}$ and $\rho_{1}$, using an optimal coherent
detector acting on all the modes. This is realized by an Helstrom
POVM (inset a). Multi-copy discrimination of two Gaussian states,
$\rho_{0}$ and $\rho_{1}$, using an incoherent detector based on a
single-mode Gaussian measurements on modes $B$ followed by an
Helstrom measurement on modes $A$ (inset b), or using an
incoherent detector based single-mode Gaussian measurements on all
the modes, whose outputs are suitably
post-processed (inset c).}%
\label{pic3}%
\end{figure}

In general, the problem can be extended to multicopy discrimination, where the
encoder prepares $M$ copies $\rho_{0}^{\otimes M}:=\rho_{0}\otimes
\cdots\otimes\rho_{0}$ or $\rho_{1}^{\otimes M}:=\rho_{1}\otimes\cdots
\otimes\rho_{1}$, with the same probability. In this generalization, we
consider a decoder that either performs an optimal global measurement
(Helstrom POVM) on all the collection of modes $A_{1},\ldots,A_{M}%
,B_{1},\ldots,B_{M}$ (see Fig.~\ref{pic3}) or a local measurement,
composed of optimal local measurements on the $A$ modes, followed
by an optimal Helstrom measurement on all the $B$ modes. This
second strategy is clearly local in terms of the $A$-$B$
bipartition and it represents an upper bound for all the
measurements based on single-mode Gaussian POVMs (see
Fig.~\ref{pic3}).

Suppose that the decoder applies an optimal global measurement. Then, for
large number of copies ($M\gg1$), the minimum error probability can be written
as%
\begin{equation}
P\approx\frac{1}{2}\exp(-\kappa M)~,
\end{equation}
where the error-probability exponent $\kappa=-\ln Q$ is provided by the
quantum Chernoff bound (QCB)~\cite{Aude,QCB}%
\begin{equation}
Q=\inf_{0\leq s\leq1}\mathrm{Tr}\left(  \rho_{0}^{s}\rho_{1}^{1-s}\right)  ~.
\label{QCB}%
\end{equation}

By contrast, if the decoder performs the local measurement (Gaussian followed
by Helstrom POVM), then the minimum error probability will have an
error-probability exponent $\kappa_{loc}\leq\kappa$. As measures of the
asymptotic gain we can consider the difference between the two error-exponents%
\[
\Delta:=\kappa-\kappa_{loc}\geq0~.
\]
As before, we are interested in studying the gain $\Delta$ in terms of the
encoded correlations $\delta D$ and $\delta C$. In this case, we are able to
prove that the gain is strictly positive and increasing in the correlations.
In particular, we have $\Delta\rightarrow\Delta_{\max}>2$ for maximal
correlations, i.e., for $\delta D\rightarrow1$ and $\delta C\rightarrow
+\infty$. Thus, the use of a global coherent measurement outperforms any local
measurements based on single-mode Gaussian detectors for discriminating the
presence or not of correlations in Gaussian states (both classical and quantum correlations).

\section{Single-copy discrimination of two-mode Gaussian states}

In our analysis we can reduce Gaussian states to zero-mean Gaussian states,
which are fully characterized by their covariance matrix (CM). This is because
displacements are local operations and therefore not related with their
correlation properties. Since discord and classical correlations are defined
as entropic quantities, they are invariant under local
unitaries~\cite{VedralRMP}. Thus, by using local Gaussian unitaries
(corresponding to symplectic transformations in the phase-space), we can
always reduce the CM of a symmetric Gaussian state $\rho$ of modes $A$ and $B$
into the normal form%
\begin{equation}
\mathbf{V}(\mu,g,g^{\prime})=\left(
\begin{array}
[c]{cccc}%
\mu &  & g & \\
& \mu &  & g^{\prime}\\
g &  & \mu & \\
& g^{\prime} &  & \mu
\end{array}
\right)  ~,
\end{equation}
where the real parameters $\mu$, $g$ and $g^{\prime}$ must satisfy a set of
bona-fide conditions~\cite{Alex}. These conditions are here equal to $\mu
\geq1$ and%
\begin{equation}
\left\vert g\right\vert <\mu,~\left\vert g^{\prime}\right\vert <\mu,~\mu
^{2}+gg^{\prime}-1\geq\mu\left\vert g+g^{\prime}\right\vert ~.
\label{bonafide}%
\end{equation}
The parameter $\mu$ quantifies the variance of the thermal noise which is
present in each bosonic mode. In fact, the two reduced states $\rho
^{A}=\mathrm{Tr}_{B}(\rho)$ and $\rho^{B}=\mathrm{Tr}_{A}(\rho)$ are identical
thermal states with mean number of photons equal to $\bar{n}=(\mu-1)/2$. The
two parameters $g$ and $g^{\prime}$ describe the correlations between the two
modes. For $g=g^{\prime}=0$ we have zero correlations ($C_{0}=D_{0}=0$) and
this is chosen to be the CM $\mathbf{V}_{0}$ of the first Gaussian state
$\rho_{0}$. The other Gaussian state $\rho_{1}$ is chosen to be separable but
with maximal correlations, both quantum and classical. Keeping fixed $\mu$,
i.e., the energy of the state, the most correlated but still separable
Gaussian state corresponds to the choice $g=g^{\prime}=\mu-1$, or equivalently
$1-\mu$. This Gaussian state $\rho_{1}$ with CM\ $\mathbf{V}_{1}%
(\mu)=\mathbf{V}(\mu,\mu-1,\mu-1)$\ has maximal classical correlations
\begin{equation}
C_{1}=h(\mu)-h\left(  \frac{3\mu-1}{\mu+1}\right)  :=\delta C(\mu)~,
\label{C1}%
\end{equation}
and maximal quantum discord~\cite{OptimalityDiscord}%
\begin{equation}
D_{1}=h(\mu)-h(2\mu-1)+h\left(  \frac{3\mu-1}{\mu+1}\right)  :=\delta D(\mu)~,
\label{D1}%
\end{equation}
where $h(x)$ is von Neumann entropy given by the formula%
\begin{equation}
h(x):=\left(  \frac{x+1}{2}\right)  \log_{2}\left(  \frac{x+1}{2}\right)
-\left(  \frac{x-1}{2}\right)  \log_{2}\left(  \frac{x-1}{2}\right)  .
\label{ent}%
\end{equation}
Our first aim is to derive the minimum error probability $P=P(\mu)$ in the
discrimination of these two Gaussian states, by performing an optimal global measurement.

\subsection{Performance of a global two-mode measurement}

Given two Gaussian states, it is generally difficult to compute the exact
performance of the optimal coherent detector, because of the trace distance
involved in the Helstrom bound. Here we resort to easier-to-compute bounds in
order to provide an estimate of the mimimum error probability. The first bound
is the QCB (single-copy formula)%
\begin{equation}
P\leq P^{+}=\frac{1}{2}\inf_{0\leq s\leq1}Q_{s}~, \label{che1}%
\end{equation}
where the s-overlap $Q_{s}$ is given by%
\begin{equation}
Q_{s}=\mathrm{Tr}(\rho_{0}^{s}\rho_{1}^{1-s})~. \label{che2}%
\end{equation}
We have a closed formula~\cite{QCB,Asym} for the s-overlap $Q_{s}$
in the case of multi-mode Gaussian states. Let us write this
formula explicitly for the case of zero-mean two-mode Gaussian
states. First introduce the two real functions%
\begin{equation}
G_{s}(x):=\frac{2^{s}}{\left(  x+1\right)  ^{s}-\left(  x-1\right)  ^{s}},~~\Lambda_{s}(x):=\frac{\left(  x+1\right)  ^{s}+\left(  x-1\right)  ^{s}%
}{\left(  x+1\right)  ^{s}-\left(  x-1\right)  ^{s}}~.
\end{equation}
Then, consider the symplectic decomposition of their CMs, $\mathbf{V}_{0}$ and
$\mathbf{V}_{1}$, according to Williamson's theorem~\cite{Willy}%
\begin{equation}
\mathbf{V}_{0}=\mathbf{S}_{0}\left(
\begin{array}
[c]{cc}%
\alpha_{-}\mathbf{I} & \\
& \alpha_{+}\mathbf{I}%
\end{array}
\right)  \mathbf{S}_{0}^{T},~~~\mathbf{V}_{1}=\mathbf{S}_{1}\left(
\begin{array}
[c]{cc}%
\beta_{-}\mathbf{I} & \\
& \beta_{+}\mathbf{I}%
\end{array}
\right)  \mathbf{S}_{1}^{T}~.
\end{equation}
Here $\alpha_{\pm}$ is the symplectic spectrum of $\mathbf{V}_{0}$
diagonalized by the symplectic matrix $\mathbf{S}_{0}$, and $\beta_{\pm}$ is
the symplectic spectrum of $\mathbf{V}_{1}$ diagonalized by the symplectic
$\mathbf{S}_{1}$~\cite{RMP}. In this case, the $s$-overlap $Q_{s}$ of the QCB
is given by the formula%
\begin{equation}
Q_{s}=\Pi_{s}\left(  \det\boldsymbol{\Sigma}_{s}\right)  ^{-1/2}
\label{sOVERLAP}%
\end{equation}
where%
\begin{equation}
\Pi_{s}:=4G_{s}(\alpha_{+})G_{s}(\alpha_{-})G_{1-s}(\beta_{+})G_{1-s}%
(\beta_{-})
\end{equation}
and
\begin{equation}
\boldsymbol{\Sigma}_{s}:=\mathbf{S}_{0}\left(
\begin{array}
[c]{cc}%
\Lambda_{s}(\alpha_{-})\mathbf{I} & \\
& \Lambda_{s}(\alpha_{+})\mathbf{I}%
\end{array}
\right)  \mathbf{S}_{0}^{T}+\mathbf{S}_{1}\left(
\begin{array}
[c]{cc}%
\Lambda_{1-s}(\beta_{-})\mathbf{I} & \\
& \Lambda_{1-s}(\beta_{+})\mathbf{I}%
\end{array}
\right)  \mathbf{S}_{1}^{T}.
\end{equation}

In order to apply the formula to the Gaussian states of our discrimination
problem, we have to derive the explicit symplectic decompositions of their
CMs. In the case of the uncorrelated thermal state $\rho_{0}$, it is trivial
to say that $\mathbf{V}_{0}$\ is already in its diagonal Williamson's form
($\mathbf{S}_{0}=\mathbf{I}$) with degenerate spectrum $\alpha_{+}=\alpha
_{-}=\mu$. For the other CM $\mathbf{V}_{1}$, it is easy to check that this is
diagonalized by a symplectic matrix of the form $\mathbf{S}_{1}=\mathbf{O}%
(\mathbf{L\oplus L)}$, where $\mathbf{O}$ is a special orthogonal matrix
(therefore symplectic) and $\mathbf{L}$ is the squeezing matrix
\begin{equation}
\mathbf{L=}\left(
\begin{array}
[c]{cc}%
\frac{1}{\sqrt{\beta}} & \\
& \sqrt{\beta}%
\end{array}
\right)  ,~\beta:=\sqrt{2\mu-1}~.
\end{equation}
The corresponding symplectic spectrum is degenerate and equal to $\beta
_{+}=\beta_{-}=\beta$ (see Appendix~\ref{AppSYM}).

The numerator in the $s$-overlap is therefore equal to%
\begin{equation}
\Pi_{s}:=4\left[  G_{s}(\mu)G_{1-s}\left(  \beta\right)  \right]  ^{2}~.
\label{piOK}%
\end{equation}
At the denominator, we have the sigma-matrix%
\begin{equation}
\boldsymbol{\Sigma}_{s}=\Lambda_{s}(\mu)(\mathbf{I\oplus I})+\Lambda
_{1-s}\left(  \beta\right)  \mathbf{O}(\mathbf{L}^{2}\mathbf{\oplus L}%
^{2}\mathbf{)O}^{T}~.
\end{equation}
Since the determinant does not change under orthogonal transformations, i.e.,
$\det\boldsymbol{\Sigma}_{s}=\det(\mathbf{O}^{T}\boldsymbol{\Sigma}%
_{s}\mathbf{O})$, we can replace $\boldsymbol{\Sigma}_{s}$ by the diagonal
matrix
\begin{align}
\boldsymbol{\Sigma}_{s}  &  =\Lambda_{s}(\mu)(\mathbf{I\oplus I}%
)+\Lambda_{1-s}\left(  \beta\right)  (\mathbf{L}^{2}\mathbf{\oplus L}%
^{2}\mathbf{)}\nonumber\\
&  =%
%TCIMACRO{\dbigoplus \limits_{i=A,B}}%
%BeginExpansion
{\displaystyle\bigoplus\limits_{i=A,B}}
%EndExpansion
\left(
\begin{array}
[c]{cc}%
\Lambda_{s}(\mu)+\frac{\Lambda_{1-s}\left(  \beta\right)  }{\beta} & \\
& \Lambda_{s}(\mu)+\beta\Lambda_{1-s}\left(  \beta\right)
\end{array}
\right)  . \label{sigmaOK}%
\end{align}
By replacing Eqs.~(\ref{piOK}) and~(\ref{sigmaOK}) into Eq.~(\ref{sOVERLAP}),
we get the analytical expression of the $s$-overlap $Q_{s}$ (not reported here
for brevity). By optimizing in the $s$ parameter as in Eq.~(\ref{che1}), we
derive the QCB as a function of $\mu$, i.e., $P^{+}=P^{+}(\mu)$. In turn, this
bound can be expressed in terms of the encoded Gaussian dicord $P^{+}%
=P^{+}(\delta D)$ and the encoded classical correlations $P^{+}=P^{+}(\delta
C)$ by using Eqs.~(\ref{C1}) and~(\ref{D1}).

\begin{figure*}[ptbh]
\vspace{-0.0cm}
\par
\begin{center}
\includegraphics[width=0.42\textwidth] {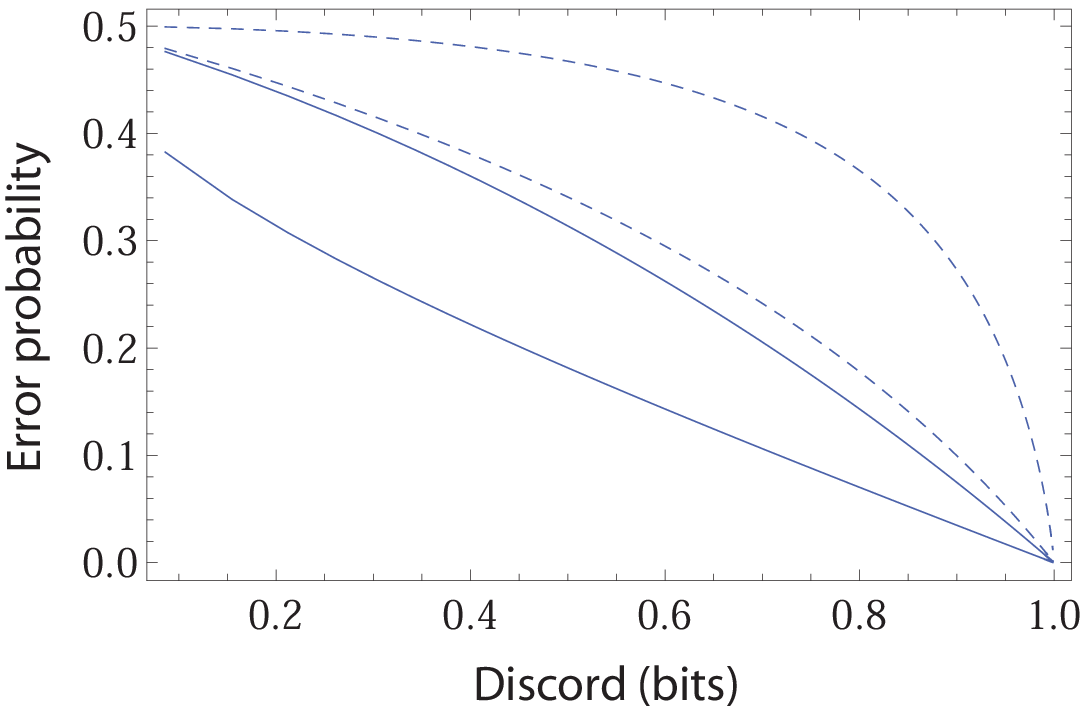}
\includegraphics[width=0.42\textwidth] {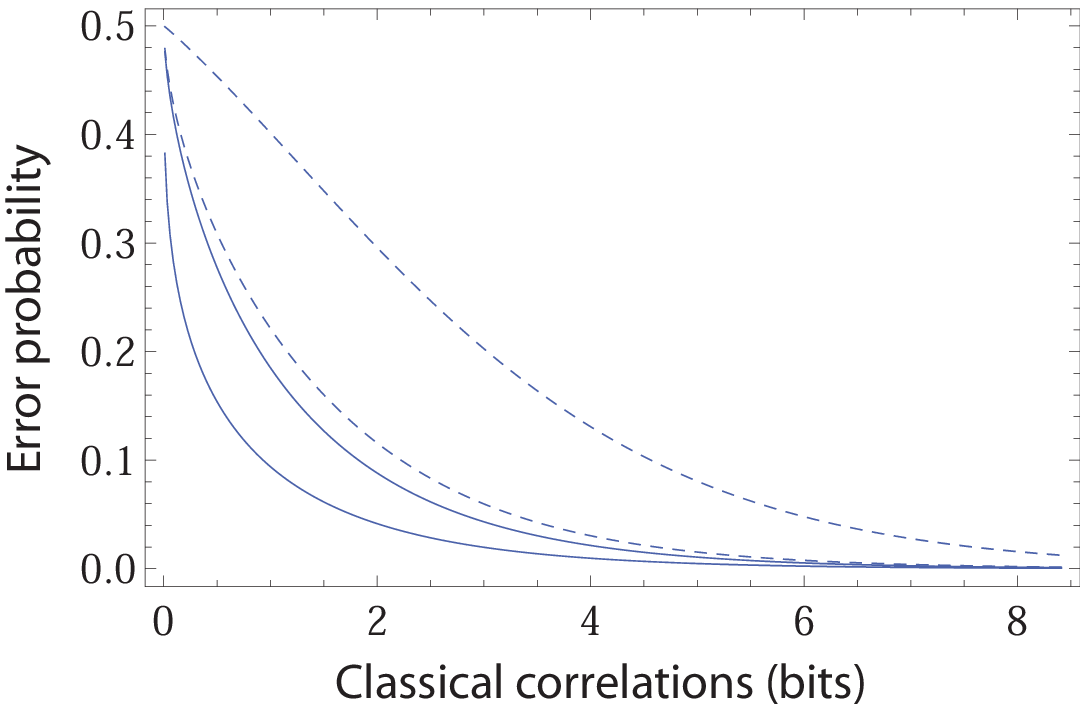}
\end{center}
\par
\vspace{-0.0cm}\caption{\textbf{Left panel}. Upper ($P^{+}$) and
lower ($P^{-}$) bounds to the error probability versus encoded
discord $\delta D$ for the optimal global detector (solid lines)
and the local detector (dashed lines). \textbf{Right panel}. The
previous bounds are plotted versus the
encoded classical correlations $\delta C$.}%
\label{compOKpic}%
\end{figure*}

Now let us derive a lower bound to the error probability. A simple bound can
be constructed using the quantum Battacharryya bound~\cite{QCB,Batta}. In
fact, we can write%
\begin{equation}
P(\mu)\geq P^{-}(\mu):=\frac{1-\sqrt{1-B^{2}}}{2}~,
\end{equation}
where%
\begin{equation}
B:=Q_{1/2}=\mathrm{Tr}(\sqrt{\rho_{0}}\sqrt{\rho_{1}})~.
\end{equation}
As before, we can express $P^{-}$ is terms of the encoded discord
$\delta D$ and classical correlations $\delta C$. The behaviour of
the two bounds $P^{+}$ and $P^{-}$ in terms of the correlations
are shown in Fig.~\ref{compOKpic}. As expected the error
probability goes to zero in the limit of maximal discord $\delta
D\rightarrow1$ and maximal classical correlations $\delta
C\rightarrow+\infty$.

\begin{figure}[ptbh]
\vspace{-0cm}
\par
\begin{center}
\includegraphics[width=0.4\textwidth] {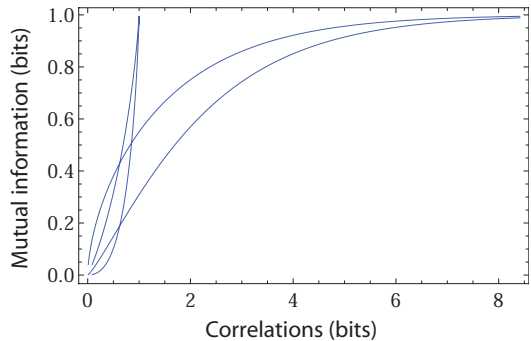}
\end{center}
\par
\vspace{-0.2cm}\caption{Upper ($I_{+}$) and lower ($I_{-}$) bounds
to the mutual information versus encoded discord (left curves) and
encoded classical
correlations (right curves)}%
\label{InfoGainOKpic}%
\end{figure}

Clearly, from the previous bounds on the error probability we can derive upper
and lower bounds for the mutual information%
\begin{equation}
I_{-}(\mu)\leq I(\mu)\leq I_{+}(\mu)~,
\end{equation}
where%
\begin{equation}
I_{+}(\mu):=1-H[P^{-}(\mu)]~,
\end{equation}
and%
\begin{equation}
I_{-}(\mu):=1-H[P^{+}(\mu)]~.
\end{equation}
The two bounds, $I_{-}$ and $I_{+}$, can be expressed in terms of $\delta C$
and $\delta D$ and are plotted in Fig.~\ref{InfoGainOKpic}%
.

\subsection{Performance of local measurements}

Here we estimate the minimum error probability which is achievable by using
local detections. As discussed in the introduction, we consider a local
Gaussian measurement followed by a local Helstrom measurement, which is an
upper bound to considering two local Gaussian detections. The most general
single-mode Gaussian POVM $\mathcal{G}$ has measurement operator
$G(\mathbf{x}):=\pi^{-1}W(\mathbf{x})\sigma W(-\mathbf{x})$, where
$\mathbf{x}\in\mathbb{R}^{2}$, $W(\mathbf{x})$ is the Weyl-displacement
operator~\cite{RMP} and $\sigma$ is a zero-mean Gaussian state with CM
\begin{equation}
\mathbf{V}_{\sigma}=\eta\mathbf{R}(\theta)\left(
\begin{array}
[c]{cc}%
\lambda & \\
& \lambda^{-1}%
\end{array}
\right)  \mathbf{R}(-\theta)~, \label{VgPOVM}%
\end{equation}
where $\eta\geq1$, $\lambda>0$ and $\mathbf{R}(\theta)$ is the rotation matrix
with angle $\theta$.

Since the encoding states are symmetric, there is no difference in where the
first measurement is performed. Without loss of generality, we assume that
mode $B$ is measured first. The reduced state of mode $B$ is a thermal state,
no matter what the encoding is ($\rho_{0}^{B}=\rho_{1}^{B}$). This means that
the outcome of the Gaussian measurement on mode $B$ does not contain any
information on the encoded two-mode state. The effect of this measurement is
only that of preparing conditional states on the other mode $A$. More
precisely, if the encoded two-mode state were $\rho_{0}$, then the conditional
state trivially coincides with the thermal reduced state $\rho_{0}^{A|B}%
=\rho_{0}^{A}$. By contrast, if the encoded state were $\rho_{1}$, then we
have the remote preparation of a Gaussian state $\rho_{1}^{A|B}(\mathbf{a})$
with conditional CM~\cite{NOTA1}%
\begin{equation}
\mathbf{V}_{A|B}=\mu\mathbf{I}-\mathbf{\tilde{V},~\tilde{V}}:=(\mu-1)^{2}%
(\mu\mathbf{I+V}_{\sigma})^{-1}~, \label{condAB}%
\end{equation}
and random displacement $\mathbf{a}$ depending on the outcome of the
measurement $\mathbf{x}$. Since the classical outcome $\mathbf{x}$ is
Gaussianly distributed, also the displacement $\mathbf{a}$\ is
Gaussianly-modulated, with a Gaussian $G_{\mathbf{\tilde{V}}}(\mathbf{a})$
having zero mean and classical CM $\mathbf{\tilde{V}}$.

Once the outcome $\mathbf{x}$ has been retrieved and therefore the remote
displacement $\mathbf{a}$ is known, the aim of the second measurement is to
distinguish between $\rho_{0}^{A|B}$ and $\rho_{1}^{A|B}(\mathbf{a})$. There
will be a corresponding Helstrom POVM which projects on the positive part of
$\gamma(\mathbf{a}):=\rho_{1}^{A|B}(\mathbf{a})-\rho_{0}^{A|B}$ with
associated error probability%
\begin{equation}
P_{loc}(\mathbf{a},\mathcal{G})=\frac{1}{2}\left[  1-T\left(  \rho_{1}%
^{A|B}(\mathbf{a}),\rho_{0}^{A|B}\right)  \right]  ~.
\end{equation}
By averaging over the random displacement, we derive the mean error
probability%
\begin{equation}
P_{loc}(\mathcal{G})=\int d\mathbf{a~}G_{\mathbf{\tilde{V}}}(\mathbf{a}%
)P_{loc}(\mathbf{a},\mathcal{G})~.
\end{equation}
Now the optimal performance is given by optimizing over all the Gaussian POVMs%
\begin{equation}
P_{loc}=\min_{\mathcal{G}}P_{loc}(\mathcal{G})~.
\end{equation}

In this case too we resort to upper and lower bounds. First we construct the
upper bound using the QCB. We have%
\[
P_{loc}(\mathbf{a},\mathcal{G})\leq P_{loc}^{+}(\mathbf{a},\mathcal{G}%
)=\frac{1}{2}\inf_{0\leq s\leq1}Q_{s}(\mathbf{a},\mathcal{G})
\]
where $Q_{s}$ must be computed here on the two Gaussian states $\rho_{0}%
^{A|B}$ (with zero mean and CM\ $\mu\mathbf{I}$) and $\rho_{1}^{A|B}%
(\mathbf{a})$, with random displacement $\mathbf{a}$ and CM $\mathbf{V}_{A|B}%
$. The latter has symplectic decomposition $\mathbf{V}_{A|B}=\nu
_{A|B}\mathbf{S}_{A|B}\mathbf{S}_{A|B}^{T}$, with $\nu_{A|B}=\sqrt
{\det\mathbf{V}_{A|B}}$ and $\mathbf{S}_{A|B}$ being a suitable symplectic
matrix. In this case the s-overlap takes the form%
\begin{equation}
Q_{s}(\mathbf{a},\mathcal{G})=\Pi_{s}\left(  \det\boldsymbol{\Sigma}%
_{s}\right)  ^{-1/2}\exp\left(  -\frac{\mathbf{a}^{T}\boldsymbol{\Sigma}%
_{s}^{-1}\mathbf{a}}{2}\right)  ~,
\end{equation}
where
\begin{align}
\Pi_{s}  &  =2G_{s}(\mu)G_{1-s}\left(  \nu_{A|B}\right)  ~,\label{pis1}\\
\boldsymbol{\Sigma}_{s}  &  =\Lambda_{s}(\mu)\mathbf{I}+\Lambda_{1-s}\left(
\nu_{A|B}\right)  \mathbf{S}_{A|B}\mathbf{S}_{A|B}^{T}~. \label{sigmas2}%
\end{align}
By averaging over the random displacement we get%
\begin{equation}
P_{loc}^{+}(\mathcal{G})=\int d\mathbf{a~}G_{\mathbf{\tilde{V}}}%
(\mathbf{a})P_{loc}^{+}(\mathbf{a},\mathcal{G})=\frac{1}{2}\inf_{0\leq s\leq
1}Q_{s}(\mathcal{G})~,
\end{equation}
where%
\begin{equation}
Q_{s}(\mathcal{G})=\frac{\Pi_{s}}{\sqrt{\det(\boldsymbol{\Sigma}%
_{s}+\mathbf{\tilde{V}})}}~. \label{QsG}%
\end{equation}

As we show in the Appendix~\ref{HetAPP}, for any $0\leq s\leq1$ we have that
$Q_{s}(\mathcal{G})$ is minimized by the heterodyne detection, which is a
Gaussian POVM with $\sigma=\left\vert 0\right\rangle \left\langle 0\right\vert
$, i.e., $\mathbf{V}_{\sigma}=\mathbf{I}$. In this case, Eq.~(\ref{condAB})
becomes%
\begin{equation}
\mathbf{V}_{A|B}=(1+\varepsilon)\mathbf{I~,~\tilde{V}}=(\mu-1-\varepsilon
)\mathbf{I} \label{casoHET}%
\end{equation}
with%
\begin{equation}
\varepsilon:=\frac{2(\mu-1)}{\mu+1}~.
\end{equation}
As we can see from Eq.~(\ref{casoHET}), by heterodyning mode $B$ we prepare
randomly-displaced thermal states on mode $A$. Furthermore, the relation
between outcome of the measurement and remote displacement is simply given by
$\mathbf{a}=(\varepsilon/\sqrt{2})\mathbf{x}$.

In the case of the heterodyne detection, the symplectic decomposition of the
conditional CM $\mathbf{V}_{A|B}$ is trivial, with eigenvalue $\nu
_{A|B}=1+\varepsilon$ and symplectic $\mathbf{S}_{A|B}=\mathbf{I}$. By
replacing these expressions in Eqs.~(\ref{pis1}) and~(\ref{sigmas2}) we get
the expressions of $\Pi_{s}$ and $\boldsymbol{\Sigma}_{s}$ to be used in
Eq.~(\ref{QsG}), where the modulation CM $\mathbf{\tilde{V}}$ is given in
Eq.~(\ref{casoHET}). As a result we get the following expression%
\begin{equation}
Q_{s}(\text{Het})=\min_{\mathcal{G}}Q_{s}(\mathcal{G})=\frac{2G_{s}%
(\mu)G_{1-s}\left(  1+\varepsilon\right)  }{\Lambda_{s}(\mu)+\Lambda
_{1-s}\left(  1+\varepsilon\right)  +\frac{(\mu-1)\varepsilon}{2}},
\end{equation}
which provides the upper bound
\begin{equation}
P_{loc}^{+}=\min_{\mathcal{G}}P_{loc}^{+}(\mathcal{G})=\frac{1}{2}\inf_{0\leq
s\leq1}Q_{s}(\text{Het})~.
\end{equation}
Thus, by minimizing in $s$ we finally get $P_{loc}^{+}=P_{loc}^{+}(\mu)$.
Using Eqs.~(\ref{C1}) and~(\ref{D1}), this bound can be expressed in terms of
encoded discord $P_{loc}^{+}=P_{loc}^{+}(\delta D)$ and classical correlations
$P_{loc}^{+}=P_{loc}^{+}(\delta C)$.

Then, we compute a lower bound by resorting to the quantum fidelity. In fact,
we can write%
\begin{equation}
P_{loc}(\mathbf{a},\mathcal{G})\geq P_{loc}^{-}(\mathbf{a},\mathcal{G}%
)=\frac{1-\sqrt{1-F(\mathbf{a},\mathcal{G})}}{2}~, \label{Ploc1}%
\end{equation}
where%
\begin{equation}
F(\mathbf{a},\mathcal{G})=\left[  \mathrm{Tr}\sqrt{\sqrt{\rho_{0}^{A|B}}%
\rho_{1}^{A|B}(\mathbf{a})\sqrt{\rho_{0}^{A|B}}}\right]  ^{2}~.
\end{equation}
As before, the bound must be averaged over the random displacement and
minimized over all the Gaussian POVMs, i.e., we must compute%
\begin{equation}
P_{loc}^{-}=\min_{\mathcal{G}}\int d\mathbf{a~}G_{\mathbf{\tilde{V}}%
}(\mathbf{a})P_{loc}^{-}(\mathbf{a},\mathcal{G})~. \label{PmenoLOC}%
\end{equation}
For two single-mode Gaussian states, the quantum fidelity can be easily
expressed in terms of the first and second order statistical
moments~\cite{Scutaru,PirFID}. Using this formula, one can prove that the
minimization in Eq.~(\ref{PmenoLOC}) is provided again by the heterodyne
detection. In this case, the optimal fidelity takes the form%
\begin{equation}
F(\mathbf{a},\text{Het})=\frac{2\exp\left[  -\frac{\varepsilon^{2}%
\mathbf{a}^{T}\mathbf{a}}{4(\mu+1+\varepsilon)}\right]  }{1+\mu(1+\varepsilon
)-2(\mu-1)\sqrt{2\mu/(\mu+1)}}.
\end{equation}
Then, using this expression in Eqs.~(\ref{Ploc1})
and~(\ref{PmenoLOC}), we get the lower bound
$P_{loc}^{-}=P_{loc}^{-}(\mu)$, which can be expressed in terms of
$\delta D$ and $\delta C$. The behaviour of the two bounds,
$P_{loc}^{-}$ and $P_{loc}^{+}$, in terms of the correlations are
shown in Fig.~\ref{compOKpic}. As we see, there is a clear
separation between the nonlocal and the local detector.

\section{Asymptotic discrimination of two-mode Gaussian states}

We now consider the case of multi-copy discrimination, where the
detector must distinguish between the two equiprobable $M$-copy
states $\rho_{0}^{\otimes M}$ or $\rho_{1}^{\otimes M}$. We
compare an optimal coherent detector, i.e., an Helstrom POVM
acting on all the modes $A_{1},\ldots,B_{M}$ with an incoherent
detector composed of single-mode Gaussian POVMs on the $A$ modes,
followed by an Helstrom POVM on the whole set of $B$ modes (see
Fig.~\ref{pic3}). This incoherent detector is clearly an upper
bound for all the measurements which are based on single-mode
Gaussian POVMs.

Here we consider the limit of large number of copies $M\gg1$. In this limit,
the minimum error probability $P$ is well-described by the QCB, whose
multi-copy formula is a simple generalization of the single-copy formula%
\begin{equation}
P^{+}=\frac{1}{2}Q^{M}~,~Q:=\inf_{0\leq s\leq1}Q_{s}~.
\end{equation}
The QCB provides the asymptotical expression of the error-probability
exponent, i.e., we can write
\begin{equation}
P\rightarrow\frac{1}{2}\exp(-\kappa M)~,
\end{equation}
using $\kappa=-\ln Q$.

\begin{figure}[ptbh]
\vspace{-0cm}
\par
\begin{center}
\includegraphics[width=0.4\textwidth] {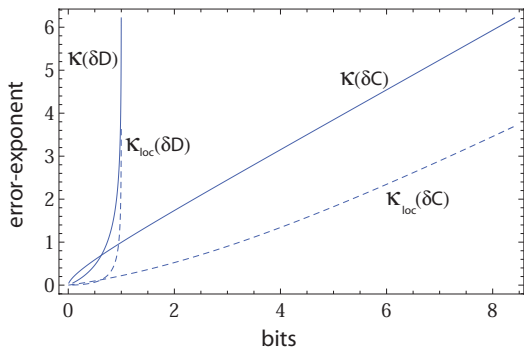}
\end{center}
\par
\vspace{-0.0cm}\caption{Error-probability exponents, $\kappa$
(coherent detector) and $\kappa_{loc}$ (incoherent detector),
plotted versus discord $\delta D$ (left curves) and classical
correlations $\delta C$ (right curves). Note the strict separation
$\kappa>\kappa_{loc}$ for any non-zero value of the
correlations.}%
\label{ExponentOKpic}%
\end{figure}

Note that we have already computed $Q$ for both the detectors. Thus, we can
easily derive $\kappa=\kappa(\mu)$ and $\kappa_{loc}=\kappa_{loc}(\mu)$ for
the coherent and incoherent detector, respectively. These quantities can then
be expressed in terms of the encoded discord $\delta D$ and classical
correlations $\delta C$. As shown in Fig.~\ref{ExponentOKpic}, there is a
strict separation $\kappa>\kappa_{loc}$ for any non-zero value of the
classical correlations.

\begin{figure}[ptbh] \vspace{-0cm}
\par
\begin{center}
\includegraphics[width=0.4\textwidth] {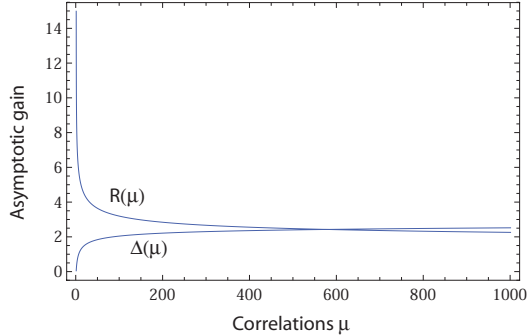}
\end{center}
\par
\vspace{-0.0cm}\caption{Asympotical gain, both quantified as a
difference $\Delta=\kappa-\kappa_{loc}$ and as a ratio
$R=\kappa/\kappa_{loc}$, versus
the encoded correlations $\mu$. Here $R$ is expressed in dB.}%
\label{ExponentGAINOKpic}%
\end{figure}

In order to better display the separation we consider the
difference between the two error-exponents
$\Delta:=\kappa-\kappa_{loc}$ which is directly plotted in terms
of the parameter $\mu$. As we can see from
Fig.~\ref{ExponentGAINOKpic}, the gain $\Delta(\mu)$ is strictly
positive and increasing in the correlations. In particular, we
have $\Delta\rightarrow \Delta_{\max}>2$ for maximal correlations,
i.e., for $\mu\rightarrow+\infty$ corresponding to $\delta
D\rightarrow1$ and $\delta C\rightarrow+\infty$. A similar
improvement can be appreciated by considering the ratio $R:=\kappa
/\kappa_{loc}\geq1$ which converges to about $2$~dB for
$\mu\rightarrow +\infty$. Thus, it is clear that the use of the
coherent detector outperforms any incoherent detector based on the
use of single-mode Gaussian measurements for the task of
discriminating the presence or not of correlations in Gaussian
states (both classical and quantum
correlations).

\section{Conclusion and discussion}

In this work, we have considered the discrimination of discord and
classical correlations in two-mode separable Gaussian states,
showing how non-local optimal detectors may retrieve strictly more
information than local Gaussian detectors. This is true both in
the setting of single- and multi-copy state discrimination.
Potential extensions of the work may involve the use of asymmetric
state discrimination, where the quantum Hoeffding
bound~\cite{Asym} and the relative entropy~\cite{relative} can be
analytically computed over Gaussian states (e.g., see Theorem 6 in
Pirandola \textit{et al.}~\cite{relative2}). It would also be
interesting to establish the actual performance of local but
non-Gaussian detectors, and also to investigate the explicit
implications for Gaussian data hiding.

\section*{Acknowledgments}
G.S. acknowledges the support of the European Commission under the Marie
Sklodowska-Curie Fellowship Progamme (EC\ grant agreement No 745727). S.P. was
supported by EPSRC via the UK quantum communications hub (EP/M013472/1). The
authors would like to thank C. Ottaviani, M. Gu and C. Weedbrook for comments
on an early draft of the manuscript.

\bigskip
%S. P. was supported by the Leverhulme Trust (`qBIO' fellowship)
%and the EPSRC via `qDATA' (EP/L011298/1) and the UK quantum
%communications hub (EP/M013472/1).

\bigskip

\appendix

\section{Symplectic decomposition of our two-mode covariance
matrix}\label{AppSYM}

Consider a CM\ in the symmetric form
\begin{equation}
\mathbf{V}=\left(
\begin{array}
[c]{cccc}%
\mu &  & g & \\
& \mu &  & g\\
g &  & \mu & \\
& g &  & \mu
\end{array}
\right)  ~,
\end{equation}
with bona-fide conditions $\mu\geq1$ and $\left\vert g\right\vert \leq\mu-1$,
coming from Eq.~(\ref{bonafide}). Since it is a real symmetric matrix, it can
be reduced to a diagonal form $\mathbf{V}=\mathbf{ODO}^{T}$ by a proper
rotation, i.e., a special orthogonal matrix $\mathbf{O}\in SO(4)$
($\mathbf{O}^{T}=\mathbf{O}^{-1}$ and $\det\mathbf{O}=+1$). In particular, we
have%
\begin{equation}
\mathbf{O}=\frac{1}{\sqrt{2}}\left(
\begin{array}
[c]{cc}%
-\mathbf{X} & \mathbf{X}\\
\mathbf{X} & \mathbf{X}%
\end{array}
\right)  ,~\mathbf{X:}=\left(
\begin{array}
[c]{cc}%
0 & 1\\
1 & 0
\end{array}
\right)  ~,
\end{equation}
and the diagonal form is composed of two degenerate eigenvalues
\begin{equation}
\mathbf{D}=\left(
\begin{array}
[c]{cc}%
(\mu-g)\mathbf{I} & \\
& (\mu+g)\mathbf{I}%
\end{array}
\right)  ~.
\end{equation}
It is easy to check that $\mathbf{O}$ is not symplectic. However, starting
from $\mathbf{O}$, it is very easy to construct a symplectic matrix
$\mathbf{S}$ such that
\begin{equation}
\mathbf{V}=\mathbf{SDS}^{T}. \label{VSDS}%
\end{equation}
In fact, it is sufficient to consider
\begin{equation}
\mathbf{S}=\mathbf{O}(\mathbf{Z}\oplus\mathbf{Z})~, \label{diagS}%
\end{equation}
where we have introduced the reflection matrix
\begin{equation}
\mathbf{Z}=\left(
\begin{array}
[c]{cc}%
1 & \\
& -1
\end{array}
\right)  ~.
\end{equation}
One can check that $\mathbf{S}$ preseves the symplectic form, i.e.,
$\mathbf{S}\boldsymbol{\Omega}\mathbf{S}^{T}=\boldsymbol{\Omega}$, with
\begin{equation}
\boldsymbol{\Omega}=\boldsymbol{\omega}\oplus\boldsymbol{\omega}%
,~\boldsymbol{\omega}:=\left(
\begin{array}
[c]{cc}%
0 & 1\\
-1 & 0
\end{array}
\right)  ~.
\end{equation}
Then, since $\mathbf{Z}^{2}=\mathbf{I}$, we have that the decomposition of
Eq.~(\ref{VSDS}) holds true. Thus, we have that $\mathbf{D}$ of
Eq.~(\ref{VSDS}) is Williamson's normal form of the CM, with symplectic
spectrum $\nu_{\pm}=\mu\pm g$, and $\mathbf{S}$ of Eq.~(\ref{diagS}) is the
diagonalizing symplectic matrix. Setting $g=\mu-1$ we retrieve the same
symplectic eigenvalues, $\nu_{-}=1$ and $\nu_{+}=2\mu-1$, given in the main text.

\section{Optimality of heterodyne}\label{HetAPP}

As we can see from Eq.~(\ref{VgPOVM}) a single-mode Gaussian POVM is
characterized by three parameters $\mathcal{G}=\mathcal{G}(\eta,\theta
,\lambda)$. In our search for the optimal POCM, we can heuristically reduce
the number of parameters from 3 to just $\lambda$ (a more rigorous proof of
this reduction can be given, but it is omitted for brevity).

Note that the Gaussian POVM on mode $B$ has an effect on mode $A$ if and only
the initial two-mode state is correlated ($\rho_{1}$), and its effect is the
preparation of a randomly-displaced Gaussian state $\rho_{1}^{A|B}%
(\mathbf{a})$. This state has to be distinguished from a thermal state
$\rho_{0}^{A}$, which is equal to averaging $\rho_{1}^{A|B}(\mathbf{a})$
averaged over the random modulation, i.e.,
\begin{equation}
\rho_{0}^{A}=\int d\mathbf{a}~G_{\tilde{V}}(\mathbf{a})\rho_{1}^{A|B}%
(\mathbf{a})~.
\end{equation}
First of all, it is intuitive to understand that the optimal Gaussian POVM is
rank-1, which corresponds to taking $\eta=1$. In fact, if we take $\eta>1$, we
see from Eq.~(\ref{condAB}) that the modulation $\mathbf{\tilde{V}}$ decreases
and, correspondingly, the conditional CM\ $\mathbf{V}_{A|B}$ increases (where
the increase/decrease must be intended here as an increase/decrease in the
symplectic eigenvalue or determinant). This clearly reduces the
distinguishability of $\rho_{1}^{A|B}(\mathbf{a})$ from the average thermal
state $\rho_{0}^{A}$.

Then, among the rank-1 Gaussian POVMs $\mathcal{G}(1,\theta,\lambda)$, no
angle $\theta$ is preferred. In fact, if $\lambda=1$ (heterodyne) we have that
any rotation in Eq.~(\ref{VgPOVM}) is equivalent to the identity. If
$\lambda\neq1$, the measurement remotely prepares a displaced thermal state
which is squeezed and rotated by $\theta$. However, since the alternative
state ($\rho_{0}^{A}$) is isotropic in phase space, there is no advantage in
preparing states which are squeezed in a particular direction.

Thus, we are left with rank-1 Gaussian POVMs of the form $\mathcal{G}%
(1,0,\lambda)$ and we want to prove that the optimal is achieved by
$\lambda=1$. To prove this, let us consider $\rho_{1}$ having a more general
CM, given by the blockform%
\begin{equation}
\mathbf{V}_{1}=\left(
\begin{array}
[c]{cc}%
\mu\mathbf{I} & g\mathbf{I}\\
g\mathbf{I} & \mu\mathbf{I}%
\end{array}
\right)  ~,
\end{equation}
with $\left\vert g\right\vert \leq\mu-1$. In this case, we derive two diagonal
matrices
\begin{equation}
\mathbf{\tilde{V}}=g^{2}\left(
\begin{array}
[c]{cc}%
\frac{1}{\lambda+\mu} & \\
& \frac{1}{\lambda^{-1}+\mu}%
\end{array}
\right)  ~, \label{VtildeAPP}%
\end{equation}
and%
\begin{equation}
\mathbf{V}_{A|B}=\mu\mathbf{I}-\mathbf{\tilde{V}}=\left(
\begin{array}
[c]{cc}%
d_{+} & \\
& d_{-}%
\end{array}
\right)  ~,
\end{equation}
with%
\begin{equation}
d_{\pm}:=\mu-\frac{g^{2}}{\lambda^{\pm1}+\mu}~.
\end{equation}
Then, we can write the symplectic decomposition $\mathbf{V}_{A|B}=\nu
_{A|B}\mathbf{SS}^{T}$, where $\nu_{A|B}=\sqrt{d_{+}d_{-}}$ and%
\begin{equation}
\mathbf{S}:=\left(
\begin{array}
[c]{cc}%
\sqrt[4]{\frac{d_{+}}{d_{-}}} & \\
& \sqrt[4]{\frac{d_{-}}{d_{+}}}%
\end{array}
\right)  ~.
\end{equation}
Next, we compute the quantities involved in the $s$-overlap, i.e.,%
\begin{align}
\Pi_{s}  &  =2G_{s}(\mu)G_{1-s}\left(  \sqrt{d_{+}d_{-}}\right)
\label{pisAPP}\\
\boldsymbol{\Sigma}_{s}  &  =\Lambda_{s}(\mu)\mathbf{I}+\Lambda_{1-s}%
(\sqrt{d_{+}d_{-}})\mathbf{SS}^{T}\text{.} \label{sigmasAPP}%
\end{align}
Thus, by replacing Eqs.~(\ref{VtildeAPP}), (\ref{pisAPP}) and~(\ref{sigmasAPP}%
) into Eq.~(\ref{QsG}), we get $Q_{s}=Q_{s}(\mu,g,\lambda)$. This expression
is analytical as well as its first derivative $\frac{\partial Q_{s}}%
{\partial\lambda}$. One can check that $\lambda=1$ is a critical point and
represents a global minimum for any allowable value of $s$, $\mu$ and $g$. In
particular, this is true for $g=\mu-1$.


\begin{thebibliography}{99}
\bibitem {NOTA1}In general, consider a two-mode Gaussian state $\rho_{AB}$
with zero mean and CM\
\[
\mathbf{V}=\left(
\begin{array}
[c]{cc}%
\mathbf{A} & \mathbf{C}\\
\mathbf{C}^{T} & \mathbf{B}%
\end{array}
\right)  ~.
\]
By performing a local Gaussian POVM ($\mathbf{V}_{\sigma}$) on mode $B$ with
outcome $\mathbf{x}$, we prepare a randomly-displaced Gaussian state on mode
$A$, with displacement $\mathbf{a}=\mathbf{a}(\mathbf{x})$ and conditional CM
$\mathbf{V}_{A|B}=\mathbf{A}-\mathbf{\tilde{V}}$ with%
\[
\mathbf{\tilde{V}}:=\mathbf{C}(\mathbf{B}+\mathbf{V}_{\sigma})^{-1}%
\mathbf{C}^{T}~.
\]
Note that $\mathbf{\tilde{V}}$ is a classical CM which describes the random
modulation of the state. In fact, summing this CM to the conditional CM, i.e.,
$\mathbf{\tilde{V}+V}_{A|B}$, we get exactly the quantum CM $\mathbf{A}$,
which describes the average state on mode $A$, i.e., the reduced state
$\rho_{A}=\mathrm{Tr}_{B}(\rho_{AB})$.
\end{thebibliography}
\end{document}